\documentclass[conference,a4paper]{IEEEtran}
\usepackage[utf8]{inputenc}
\usepackage{amsmath,amsfonts}
\usepackage{amssymb}
\usepackage{cite}
\newtheorem{lemma}{Lemma}

\newtheorem{example}{Example}
\newtheorem{corollary}{Corollary}
\newtheorem{theorem}{Theorem}
\newtheorem{proposition}{Proposition}
\newtheorem{definition}{Definition}
\newcommand{\expectation}{\ensuremath{\mathbb{E}}}
\newcommand{\Expt}{\expectation}
\def\fF{\mathbb{F}}

\def\cY{\mathcal{Y}}

\newcommand{\probability}{\ensuremath{\mathbb{P}}}
\newcommand{\Prob}{\probability}

\title{Universal Polar Decoding with Channel Knowledge at the Encoder}
\author{
   \IEEEauthorblockN{Mine Alsan}
   \IEEEauthorblockA{Information Theory Laboratory\\
        Ecole Polytechnique F\'ed\'erale de Lausanne\\
        CH-1015 Lausanne, Switzerland\\
        Email: mine.alsan@epfl.ch}
}
\begin{document}
\maketitle

\begin{abstract}
Polar coding over a class of binary discrete memoryless
channels with channel knowledge at the encoder is studied. 
It is shown that polar codes achieve the capacity of convex and one-sided classes of symmetric channels.  
\end{abstract}

\section{Introduction}
The study of the channel polarization phenomenon led by Ar{\i}kan reveals the design of a polar code is a channel specific task \cite{1669570}. 
A central stage in the design of these class of codes invented by Ar{\i}kan is the construction of the information set.
While designing a polar code, the method of channel combining and splitting is used to transform independent copies of a given binary discrete memoryless channel (B-DMC) into new synthetic 
channels by applying recursively the basic polarization transformations until the obtained channels are sufficiently polarized. 
Then, the information set is specified by selecting the synthetic channels' indices that are good for uncoded transmission, such as those having high mutual information or low Bhattacharyya parameters. In fact, the splitting method makes the synthesized channels suitable for a successive cancellation decoding procedure
and the main idea behind the construction is to select the information set such that the overall error probability of the decoding procedure is small. 
Once the information set is constructed, see \cite{6557004} for an efficient implementation method, 
the encoder and decoder operate with the common knowledge of the information set, knowing over which indices to transmit and decode data and over which to use priorly fixed bits. This creates a dependence of both of these system components on the communication channel. In scenarios where the channel is unknown or only a partial knowledge exists computing the information set becomes 
a challenging problem. Two exceptions are known to lead to an ordering of the information sets \cite{1669570}: Any binary erasure channel (BEC) provides good indices for all other B-DMCs
having smaller Bhattacharyya parameters, and any channel which is degraded with respect to another B-DMC provides good indices for the upgraded channel. 
So, the availability of some partial knowledge seems to help at least for constructing an information set. 

However, a subtle point is usually left aside: 
besides the information set, the polar successive cancellation decoder (SCD) requires the exact channel knowledge to function. 
In case the channel is unknown, the successive cancellation decoder will employ a possibly mismatched channel which might cause loss in the performance.
Therefore, even in the mentioned examples where an information set can be constructed without perfectly knowing the channel, 
the mismatch in the decoder has to be also taken into account. Indeed, when defining the information set, 
the parameter used in the definition of the information set has to be adapted carefully to the context of communication keeping in mind that the transmitted data need to be decoded reliably using the available decoding metric. 
This is the motivation behind the recent study in \cite{MineITW13} where we have investigated the performance of mismatched polar codes over an unknown channel and we have derived a lower bound on achievable rates by mismatched polar codes.  

Furthermore, even in the cases where both the encoder and decoder know the communication channel, practical design considerations might direct the designer to choose a mismatched decoding metric for the decoding procedure. The next example, which illustrates this scenario, further motivates the study of mismatched decoding of polar codes.
\begin{example}\label{ex::Tel}
Consider, for example, where a high signal to noise ratio channel is used with a large constellation with points indexed by k-bit symbols $s(0...0),...,s(1...1)$, and the receiver is interested in recovering the 1st of these k bits.  Then, the true likelihood ratio requires the computation of the sums
\begin{equation}
  \sum_{b_2,...,b_k} W(y|s(0,b_2,...,b_k)\quad \& \quad
  \sum_{b_2,...,b_k} W(y|s(1,b_2,...,b_k),
\end{equation}
each containing an exponential number of terms (in k).  The receiver hardware may not permit such computations, and consequently the decoder designer may be forced to use a simpler metric $V(y|0)/V(y|1)$ which approximates the true one.  In such cases even when the receiver is informed of the true channel $W$, the decoding operation proceeds on the basis of a mismatched channel $V$. 
\end{example}

In his thesis Korada  \cite{4461THESES} studies the compound capacity of polar codes and shows that in general polar codes do not achieve the compound capacity of a class of symmetric channels \cite{articlecompound}. 
The underlying assumption throughout the analysis is the availability of the actual communication channel to the decoder. 
This transforms the problem into finding an information set providing suitable indices for communication over any channel in the class. 
As the lack of knowledge cannot possibly reverse the end result, the assumptions simplify the theoretical analysis for drawing the general conclusion. On the other hand, the following question has not been addressed so far: what rates could be achieved in the compound setting when there is a mismatch in decoding?
Both from a conceptual and practical points of view, the effects of the decoder mismatch on the achievable rates over a class of channels is an important problem to be studied. 

This paper aims to provide a partial characterization of such mismatch effects over a compound set of channels. In particular, we study universal polar decoding over a class of B-DMCs with channel knowledge at the encoder. The achievability results stated in \cite{MineITW13} are the starting point of this paper.  We first extend the result stated in \cite[Theorem 1]{MineITW13} for mismatched channels which are symmetric under the same permutation to general B-DMCs.  
The extended result is given in Theorem \ref{thm::achievability}. Then, we show in Theorem \ref{thm::compound} that polar codes do achieve the compound capacity 
of convex sets of symmetric channels and more generally of one sided-sets of symmetric channels which were introduced in \cite{abbeonesided}.

The problem of universal polar decoding over a class of B-DMCs in scenarios where the encoder knows the actual communication channel can be seen as the complement of the problem of universal polar encoding with channel knowledge at the decoder in which 
the designer needs to find a single polar encoder, i.e. an appropriate information set, such that the transmitted codewords can be reliably communicated 
over any channel in the class when decoded at the receiver side by using the appropriate successive cancellation decoder adapted to the communication channel. 
Here we remove the assumption that the decoder uses the matched decoding metric during the decision procedure. Hence, the designer is required 
to design a single successive cancellation decoder which could be used for reliably decoding over any channel in the class while the encoder and decoder need to adapt 
the information set according to the actual communication channel and the channel used in the design of the successive cancellation decoder. At this point, a comment is in order to avoid confusions. Note that for polar codes, the adjective `universal' is often used to refer to the problem of universal polar encoding with channel knowledge at the decoder. Yet, `universality' as studied by Blackwell et. al in \cite{articlecompound}, or as studied in \cite{720546} impose stronger robustness than both of the described complementary problems.\\
 
The rest of this paper is organized as follows. Section \ref{sec::pre} gives the preliminary definitions and introduces the key results of \cite{MineITW13}. Then, these results are extended in Section \ref{sec::res}, and the main results of this paper are stated in Theorem \ref{thm::achievability} and Theorem \ref{thm::compound}. Finally, Section \ref{sec::dis} concludes with some discussions on the code construction.

\section{Preliminaries}\label{sec::pre}
The compound capacity of a class of channels is given by~\cite{articlecompound}
\begin{equation}\label{eq:c}
C(\mathcal{W}) = \displaystyle\max_{Q(x)} \inf_{W \in \mathcal{W}} I(Q, W),
\end{equation}
where $\mathcal{W}$ represents the class of channels, $Q(x)$ is the input distribution, and $I(Q, W)$ is the corresponding mutual information
between the input and output of the channel
\begin{equation}
I(Q, W) = \displaystyle\sum_{y}\sum_{x\in\fF_2}Q(x)W(y|x)\log{\frac{W(y|x)}{\displaystyle\sum_{x'}Q(x')W(y|x')}}.
\end{equation}

For a symmetric class of channels, the uniform input distribution corresponds to the maximizing input distribution in \eqref{eq:c}. The compound capacity expression for a class of symmetric channels becomes 
\begin{equation}\label{eq:compound}
\mathcal{I}(\mathcal{W}) = \displaystyle\min_{W \in \mathcal{W}} I(W),
\end{equation}
where $I(W)\triangleq I(\textit{uniform}, W)$ is the symmetric capacity of the channel.
Restricting the analysis to the uniform input distribution, we can also define the `symmetric compound capacity' of a class of channels by the expression in \eqref{eq:compound}. \\

Let $W$ be a B-DMC; $W(y|x)$ denotes the transition probabilities where the input $x$ is $\{0, 1\}$ valued and the output $y$ takes values from an arbitrary discrete alphabet $\cY$. 
We abuse the notation to define $W(x|y)$ as the inputs' posterior probabilities given the output, and we denote the difference by $\Delta_{W}(y) = W(0|y) - W(1|y)$. 
Finally, let $q_W(y) = \left(W(y|0) + W(y|1)\right)/2$ denote the output distribution of the channel $W$ when the inputs are uniformly distributed. 
For the rest of this paper, we will assume the channel inputs are uniformly distributed.

The basic polarization transformations synthesize two new binary input channels $W^-:\fF_2\to\cY^2$ and $W^+:\fF_2\to\cY^2\times\fF_2$ from two independent copies of $W$. 
Their transition probabilities are given by \cite{1669570}
\begin{eqnarray}
&&W^-(y_1y_2|u_1)=\sum_{u_2\in\fF_2}\displaystyle\frac{1}{2} W(y_1|u_1\oplus u_2)W(y_2|u_2), \IEEEeqnarraynumspace\\
&&W^+(y_1y_2u_1|u_2)=\displaystyle\frac{1}{2} W(y_1|u_1\oplus u_2)W(y_2|u_2).
\end{eqnarray}
The recursive application of the basic polarization transformations synthesizes at level $n=1, 2, \ldots$ the channels  
$W_{N}^{(i)}\colon \mathcal{X}\to\mathcal{Y}\times\mathcal{X}^{i-1}$, for $N=2^n$, and for $i=1, \ldots, N$, with transition probabilities given in \cite[Equation (5)]{1669570}.

The polarization process for a given channel $W$ is defined as the random sequence of channels $\{W_{n}\}$ such that $W_{0} = W$, and for $n \geq 0$ \cite{5205856}
\begin{equation}
 W_{n+1} = \left\lbrace \begin{array}{lll}
                           W_{n}^{-}  & \hbox{if} \hspace{2mm} B_{n} = 0\\ W_{n}^{+} & \hbox{if} \hspace{2mm} B_{n} = 1
                          \end{array}
\right.,  
\end{equation}
where $B_1, \dots, B_{n}$ is a random i.i.d. sequence drawn according to a Bernoulli distribution with probabilities equal to $1/2$.

Given two B-DMCs $W$ and $V$, we denote
\begin{equation}
I(W, V) = \displaystyle\sum_{y}\sum_{x\in\{0,1\}}\frac{1}{2}W(y|x) \log{\frac{V(y|x)}{q_V(y)}}.
\end{equation}
It was shown in \cite{MineITW13} that the average error probability $P_{\textnormal{e}}(W, V)$ of a single use of a B-DMC $W$ to transmit a $0$ or a $1$ 
when maximum likelihood decoding with respect to a possibly mismatched B-DMC $V\colon \mathcal{X}\to\mathcal{Y}$ is used as a decoding metric is upper bounded by
\begin{equation}\label{eq::Pe_UB}
P_{\textnormal{e}}(W, V) \leq 1 - I(W, V).
\end{equation}
Note that the parameter $I(W, V)$ also appears in Fischer~\cite{Fischer} as the generalized mutual information evaluated under the uniform input distribution. 
The following lemma from \cite[Lemma 3]{MineITW13} provides a useful representation for $I(W, V)$.
\begin{lemma}\cite{MineITW13}\label{lem::mismatch_I}
\begin{IEEEeqnarray}{rCl}\label{eq::mismatch_I}
I(W, V) & = & \frac{1}{2}\displaystyle\sum_{y} W(y|0) \log\left(1 + \Delta_{V}(y)\right) \nonumber\\
&& +\: \frac{1}{2}\displaystyle\sum_{y} W(y|1) \log\left(1 - \Delta_{V}(y)\right).
\end{IEEEeqnarray}
\end{lemma}

We define the process $I(W_n, V_n)$ associated to the polarization process of
both channels. Note that $I(W_n) \triangleq I(W_n, W_n)$ corresponds to the symmetric capacity process of the channel. 
The following proposition is one of the key results derived in \cite{MineITW13}. 

\begin{proposition}\cite{MineITW13}\label{prop::I_n_martingale}
The process $I_n(W, V)$ is a bounded martingale such that $I_n(W, V)\leq 1$ holds for each $n\geq 0$, the process converges a.s., and 
$\Expt[I_\infty(W, V)] \geq I(W, V)$ holds.
\end{proposition}

Consider the random variable $\Delta_{V}(Y)$ whose distribution depends on the distribution of the outputs of the channel: if the inputs are transmitted over the channel $W$, than the distribution of $\Delta_{V}(Y)$ will be determined by $q_W(y)$. The following is another important result obtained in \cite{MineITW13}.

\begin{proposition}\cite{MineITW13}\label{prop::delta_convergence}
Let $W$ and $V$ be symmetric channels under the same permutation. Then, the process $\Expt\left[\sqrt{|\Delta_{V_n}|}\right]$ is a bounded supermartingale which converges a.s. under $q_{W_n}$ to a limiting random variable which is $\{0, 1\}$ valued. 
\end{proposition}
In Section \ref{sec::res}, the above proposition will be extended to general B-DMCs $W$ and $V$. \\

The last preliminary we introduce is concerned with classes of channels satisfying the following condition: 
\begin{equation}\label{eq::cond} 
I(W, V) \geq I(V), \quad \forall \hspace{1mm} W\in\mathcal{W}.
\end{equation}
These classes of channels will play an important role in Theorem \ref{thm::compound}. The next proposition states a result from \cite{csiszarinformation} which shows that convex sets of channels satisfy this condition. 
\begin{proposition} \cite{csiszarinformation}\label{prop:ineq2}
Given  a set of channels $\mathcal{W}$, and $\mathcal{\bar{W}}$ its convex closure, 
if $V = \displaystyle\arg\min_{W\in\mathcal{\bar{W}}}I(W)$, then
\begin{equation}
 I(\alpha W+(1-\alpha)V) \geq I(V), \quad \forall \alpha\in[0, 1], \forall\hspace{1mm} W\in\mathcal{W},
\end{equation}
which in turns implies the condition in \eqref{eq::cond}.
\end{proposition}
An example of a convex set of channels is the class of binary symmetric channels with crossover probabilities within an interval.

Note that convexity is a sufficient but not necessary condition for \eqref{eq::cond} to hold. One-sided sets defined in Abbe et. al \cite{abbeonesided} generalize this idea. The following definition is adapted from \cite[Definition 3]{abbeonesided}.

\begin{definition}\label{def::one_sided}\cite{abbeonesided}
A set $\mathcal{W}$ is called one sided with respect to the uniform input distribution
if the following minimizer is unique:
\begin{equation}
V = \displaystyle\arg\min_{W\in\mathsf{cl}(\mathcal{W})}I(W)
\end{equation}
where $\mathsf{cl}(\mathcal{W})$ is the closure of $\mathcal{W}$, and if the condition in \eqref{eq::cond} holds. 
(Though this condition is stated as a divergence inequality in \cite[Equation 8]{abbeonesided}, one can easily check it is equivalent to \eqref{eq::cond} in this case. )
\end{definition}
Moreover, by \cite[Lemma 4]{abbeonesided} convex sets are one-sided and there exist one-sided sets that are not convex.

\section{Results}\label{sec::res}
We first extend Proposition 1 to channels that are not necessarily symmetric. We first need the following lemma.
\begin{lemma}\label{lem::delta_plus}
Suppose $\Delta_1$, $\Delta_2$ are independent $[-1,1]$ valued
random variables with $\Expt\left[\sqrt{|\Delta_i|}\right]=\mu_i$.  Then
\begin{equation}
\Expt\Biggl[\sqrt{\biggl|\frac{\Delta_1+ \Delta_2}{1+\Delta_1\Delta_2}\biggr|}\Biggr]
\leq \mu_1+\mu_2-\mu_1\mu_2.
\end{equation}
and
\begin{equation}
\Expt\Biggl[\sqrt{\biggl|\frac{\Delta_1- \Delta_2}{1-\Delta_1\Delta_2}\biggr|}\Biggr]
\leq \mu_1+\mu_2-\mu_1\mu_2.
\end{equation}
\end{lemma}
\begin{IEEEproof}
By \cite[Lemma 6]{MineITW13}, we have
\begin{equation}
\sqrt{\biggl|\frac{\Delta_1+\Delta_2}{1+\Delta_1\Delta_2}\biggr|}
\leq \sqrt{|\Delta_1|}+\sqrt{|\Delta_2|}-\sqrt{|\Delta_1|}\sqrt{|\Delta_2|},
\end{equation}
\begin{equation}
\sqrt{\biggl|\frac{\Delta_1-\Delta_2}{1-\Delta_1\Delta_2}\biggr|}
\leq \sqrt{|\Delta_1|}+\sqrt{|\Delta_2|}-\sqrt{|\Delta_1|}\sqrt{|\Delta_2|}.
\end{equation}
The lemma follows by taking expectations of both sides and noting the
independence of $\Delta_1$ and $\Delta_2$.
\end{IEEEproof} 

\begin{proposition}\label{prop::delta_convergence_2}
Let $W$ and $V$ be B-DMCs. Then, the process $\Expt\left[\sqrt{|\Delta_{V_n}|}\right]$ is a bounded supermartingale which converges a.s. under $q_{W_n}$ to a limiting random variable which is $\{0, 1\}$ valued. 
\end{proposition}
\begin{IEEEproof} 
We will prove that 
\begin{multline}
 \Expt\left[\sqrt{|\Delta_{V^-}(Y_1Y_2)|}\right] + \Expt\left[\sqrt{|\Delta_{V^+}(Y_1Y_2 U_1)|}\right] \\
 \leq 2\Expt\left[\sqrt{|\Delta_{V}(Y)|}\right]
\end{multline}
holds after a single step. The general result showing the process is a supermartingale will follow by the recursive structure. First we make the following simple observations
\begin{align}
\label{eq::d_minus}\Delta_{V^{-}}(y_1y_2) &= \Delta_{V}(y_1)\Delta_{V}(y_2), \\
\label{eq::d_plus}\Delta_{V^{+}}(y_1y_2u_1) &= \displaystyle\frac{\Delta_{V}(y_1) + (-1)^{u_1} \Delta_{V}(y_2)}{1 + (-1)^{u_1}\Delta_{V}(y_1)\Delta_{V}(y_2)},
\end{align}
Using \eqref{eq::d_minus} and \eqref{eq::d_plus}, we get 
\begin{equation}
\Expt\left[\sqrt{|\Delta_{V^-}(Y_1Y_2)|}\right] = \Expt\left[\sqrt{|\Delta_{V}(Y)|}\right]^2,
\end{equation}
and
\begin{multline}
 \Expt\left[\sqrt{\left|\Delta_{V^+}(Y_1Y_2 U_1)\right|}\right] \\
  \leq 2\Expt\left[\sqrt{|\Delta_{V}(Y)|}\right] - \Expt\left[\sqrt{|\Delta_{V}(Y)|}\right]^2, 
\end{multline}
where the last inequality follows by Lemma \ref{lem::delta_plus}.
As $\sqrt{\left|\Delta_{V_n}\right|}\in[0, 1]$, this proves the process is a bounded supermartingale and converges a.s. under $q_{W_n}$. 
One can prove the convergence is to $\{0, 1\}$ 
using the squaring property of the minus transformation in a similar fashion as in the proof of \cite[Proposition 9]{1669570}.
\end{IEEEproof}

\begin{corollary}\label{cor::I_convergence}
By Proposition \ref{prop::delta_convergence_2}, the distribution of $\Delta_{V_\infty}$ should be supported at $\{-1, 0, 1\}$. So by Lemma \ref{lem::mismatch_I}, $I_n(W, V)$ can only converge to the values $\{-\infty, 0, 1\}$.
\end{corollary}

Now, we give the main results in the next two theorems. 
\begin{theorem}\label{thm::achievability}
Let $W$ and $V$ be two B-DMCs such that $I(W, V) > -\infty$. Then, $I_\infty(W, V)$ is $\{0, 1\}$ valued with
\begin{equation}\label{eq::pos_rate}
\Prob[I_\infty(W, V)\to 1] \geq I(W, V).
\end{equation}
Moreover, the speed of convergence is $O(2^{-\sqrt{N}})$.
\end{theorem}
\begin{IEEEproof}
By Corollary \ref{cor::I_convergence}, we know $I_n(W, V)$ can only converge to the values $\{-\infty, 0, 1\}$ when $\Expt\left[\sqrt{|\Delta_{V_n}|}\right]$ converges.
On the other hand, by Proposition \ref{prop::I_n_martingale}, we also know that $\Expt[I_\infty(W, V)] \geq I(W, V)$. As $I(W, V) > -\infty$, 
$I_\infty(W, V)$ can only be $\{0, 1\}$ valued and \eqref{eq::pos_rate} must hold. 
The speed of convergence result follows by \cite[Theorem 1]{5205856} as 
the conditions (z.1), (z.2), (z.3) in \cite{5205856} hold taking $Z_n = \Expt\left[\sqrt{|\Delta_{V_n}|}\right]$ and $I_0 = I(W, V)$.
\end{IEEEproof}

\begin{theorem}\label{thm::compound}
Given a class of one-sided symmetric channels $\mathcal{W}$, consider the polar successive cancellation decoder using the mismatched decoding rule for the channel $V = \displaystyle\arg\min_{W\in\mathsf{cl}(\mathcal{W})}I(W)$, and the class of polar codes 
with the information sets $\mathcal{A}_{N} ^{\epsilon}(W, V) = \{i=1, \ldots, N: I(W_N^{(i)}, V_N^{(i)})\geq 1-\epsilon\}$, where $W\in\mathcal{W}$, $N = 2^n$ with $n= 1, 2, \ldots$ is the blocklength, and $\epsilon>0$ is a desired threshold. Then, for any $R < I(V)$, one can select $\epsilon \sim O(2^{-\sqrt{N}})$ and construct for all $W\in\mathcal{W}$ the information sets $\mathcal{A}_{N} ^{\epsilon}(W, V)$ of size at least as large as $NR$. Moreover, the resulting decoding error probability over any channel $W\in\mathcal{W}$ of the corresponding polar code can be made arbitrarily small by taking $N\to\infty$. In that sense, the polar SCD is universally compound capacity achieving over one-sided sets of symmetric channels.
\end{theorem}
\begin{IEEEproof}
By the speed of convergence result stated in Theorem \ref{thm::achievability}, the claim on the construction of $\mathcal{A}_{N} ^{\epsilon}(W, V)$ follows. The mismatched decoding error probability over the channel $W\in\mathcal{W}$ of a polar code with information set $\mathcal{A}_{N} ^{\epsilon}(W, V)$, for a given $N$ and $\epsilon > 0$, and using a mismatched SCD operating with the parameters of the channel $V$, will be upper bounded by
\begin{equation}
P_{\textnormal{e}, \mathsf{SCD}}(W, V) \leq \displaystyle\sum_{i\in\mathcal{A}_{N} ^{\epsilon}(W, V)} 1 - I_{N}^{(i)}(W, V),
\end{equation}
where the upper bound can be derived using the union bound and the upper bound in \eqref{eq::Pe_UB}. Taking $N\to\infty$, the chosen $\epsilon\to0$, and we get $P_{\textnormal{e}, \mathsf{SCD}}(W, V)\to 0$ by Theorem \ref{thm::achievability}. The one-sidedness  of $\mathcal{W}$ ensures that for $V$ chosen as in the hypothesis of the theorem, the relationship $I(W, V) \geq I(V)$ holds for any $W\in\mathcal{W}$. We conclude the described polar codes can achieve a rate of at least $I(V)$ over any $W\in\mathcal{W}$. 
\end{IEEEproof}

\section{Discussions}\label{sec::dis}

Finally, we discuss code construction methods for the different communication scenarios. We first propose a code construction method using the original polar code construction idea of Ar{\i}kan~\cite{1669570} in an `online' fashion to handle the scenario in which the decoder does not know the channel and can feedback. The method is based on the estimation of the parameters $I(W_N^{(i)}, V_N^{(i)})$ by a Monte Carlo approach. For that purpose, the encoder needs to perform multiple transmissions of an input. Then, the decoder must compute an estimate of the parameters by averaging. Once the information set is constructed, the decoder shall reveal this information to the encoder by feedback. 
Observe that this construction method does not even require the encoder to know the 
actual communication channel. 

In scenarios where both the encoder and decoder know the communication channel, yet mismatched decoding is performed as in Example \ref{ex::Tel}, the estimation of the parameters $I(W_N^{(i)}, V_N^{(i)})$ can be carried `offline' at both the encoder and decoder sides as opposed to the `online' computations required by the previous scenario. Moreover,
we believe computationally more efficient alternatives can be proposed by extending the efficient code construction method proposed in \cite{6557004} to the mismatched case. 
 
\section*{Acknowledgment}
The author would like to thank Emre Telatar for providing Example \ref{ex::Tel}.
This work was supported by Swiss National Science Foundation under grant number 200021-125347/1.

\bibliographystyle{IEEEtran}
\bibliography{ref}
\end{document}